\documentclass[acmsmall]{acmart}

\AtBeginDocument{%
  }

\usepackage{algorithm}
\usepackage[noend]{algpseudocode}
\usepackage{float}
\usepackage{amsmath}
\usepackage{graphicx}
\usepackage[most]{tcolorbox}
\usepackage{booktabs} %
\usepackage{seqsplit}
\usepackage{svg}
\usepackage{overpic}
\usepackage{multirow} %
\usepackage{threeparttable}
\usepackage{diagbox}
\usepackage{xspace}
\usepackage{xcolor}
\usepackage{enumitem}
\usepackage{graphicx,subcaption}
\usepackage{cleveref}

\usepackage{listings, xcolor}

\definecolor{verylightgray}{rgb}{.97,.97,.97}
\definecolor{codegreen}{rgb}{0,0.55,0}

\lstdefinelanguage{Solidity}{
	keywords=[1]{anonymous, assembly, assert, balance, break, call, callcode, case, catch, class, constant, continue, constructor, contract, debugger, default, delegatecall, delete, do, else, emit, event, experimental, export, external, false, finally, for, function, gas, if, implements, import, in, indexed, instanceof, interface, internal, is, length, library, log0, log1, log2, log3, log4, memory, modifier, new, payable, pragma, private, protected, public, pure, push, require, return, returns, revert, selfdestruct, send, solidity, storage, struct, suicide, super, switch, then, this, throw, transfer, true, try, typeof, using, value, view, while, with, addmod, ecrecover, keccak256, mulmod, ripemd160, sha256, sha3}, %
	keywordstyle=[1]\color{blue}\bfseries,
	keywords=[2]{address, bool, byte, bytes, bytes1, bytes2, bytes3, bytes4, bytes5, bytes6, bytes7, bytes8, bytes9, bytes10, bytes11, bytes12, bytes13, bytes14, bytes15, bytes16, bytes17, bytes18, bytes19, bytes20, bytes21, bytes22, bytes23, bytes24, bytes25, bytes26, bytes27, bytes28, bytes29, bytes30, bytes31, bytes32, enum, int, int8, int16, int24, int32, int40, int48, int56, int64, int72, int80, int88, int96, int104, int112, int120, int128, int136, int144, int152, int160, int168, int176, int184, int192, int200, int208, int216, int224, int232, int240, int248, int256, mapping, string, uint, uint8, uint16, uint24, uint32, uint40, uint48, uint56, uint64, uint72, uint80, uint88, uint96, uint104, uint112, uint120, uint128, uint136, uint144, uint152, uint160, uint168, uint176, uint184, uint192, uint200, uint208, uint216, uint224, uint232, uint240, uint248, uint256, var, void, ether, finney, szabo, wei, days, hours, minutes, seconds, weeks, years},	%
	keywordstyle=[2]\color{teal}\bfseries,
	keywords=[3]{block, blockhash, coinbase, difficulty, gaslimit, number, timestamp, msg, data, gas, sender, sig, value, now, tx, gasprice, origin},	%
	keywordstyle=[3]\color{violet}\bfseries,
	identifierstyle=\color{black},
	sensitive=false,
	comment=[l]{//},
	morecomment=[s]{/*}{*/},
	commentstyle=\color{codegreen}\ttfamily,
	stringstyle=\color{red}\ttfamily,
	morestring=[b]',
	morestring=[b]"
}

\usepackage{tikz}
\setcopyright{acmlicensed}
\copyrightyear{2018}
\acmYear{2018}
\acmDOI{XXXXXXX.XXXXXXX}

\acmJournal{JACM}
\acmVolume{37}
\acmNumber{4}
\acmArticle{111}
\acmMonth{8}

\newcommand{\tool}{\textsc{LogicScan}\xspace}
\newcommand{\answer}[2]{
  \begin{tcolorbox}[enhanced, left=3mm,right=3mm,
    colback=gray!10, colframe=gray!80, boxrule=0pt,
    borderline west={4pt}{0pt}{gray!90},
    breakable
    ]
    \textbf{Answer to RQ#1:} #2
    \end{tcolorbox}
}

\begin{document}

\title{LogicScan: An LLM-driven Framework for Detecting Business Logic Vulnerabilities in Smart Contracts}

\author{Jiaqi Gao}
\affiliation{%
  \institution{School of Cyberspace Science and Technology, Beijing Institute of Technology}
  \city{Beijing}
  \country{China}
}
\email{judgegao06@gmail.com}

\author{Zijian Zhang}
\authornote{Corresponding author.}
\affiliation{%
  \institution{School of Cyberspace Science and Technology, Beijing Institute of Technology}
  \city{Beijing}
  \country{China}
}
\email{zhangzijian@bit.edu.cn}

\author{Yuqiang Sun}
\authornote{Corresponding author.}
\affiliation{%
  \institution{Nanyang Technological University}
  \country{Singapore}
}
\email{yuqiang.sun@ntu.edu.sg}

\author{Ye Liu}
\affiliation{%
  \institution{School of Cyberspace Science and Technology, Beijing Institute of Technology}
  \city{Beijing}
  \country{China}
}
\email{liuye5@live.com}

\author{Chengwei Liu}
\affiliation{%
  \institution{College of Cryptology and Cyber Science, Nankai University}
  \city{Tianjin}
  \country{China}
}
\email{chengwei.liu@ntu.edu.sg}

\author{Han Liu}
\affiliation{%
  \institution{College of Cryptology and Cyber Science, Nankai University}
  \city{Tianjin}
  \country{China}
}
\email{liuhan@ust.hk}

\author{Yi Li}
\affiliation{%
  \institution{Nanyang Technological University}
  \country{Singapore}
}
\email{yi_li@ntu.edu.sg}

\author{Yang Liu}
\affiliation{%
  \institution{Nanyang Technological University}
  \country{Singapore}
}
\email{yangliu@ntu.edu.sg}

\renewcommand{\shortauthors}{JIAQI et al.}

\begin{abstract}
Business logic vulnerabilities have become one of the most damaging yet least understood classes of smart contract vulnerabilities. Unlike traditional bugs such as reentrancy or arithmetic errors, these vulnerabilities arise from missing or incorrectly enforced business invariants and are tightly coupled with protocol semantics. Existing static analysis techniques struggle to capture such high-level logic, while recent large language model based approaches often suffer from unstable outputs and low accuracy due to hallucination and limited verification.

In this paper, we propose \tool, an automated contrastive auditing framework for detecting business logic vulnerabilities in smart contracts. The key insight behind LogicScan is that mature, widely deployed on-chain protocols implicitly encode well-tested and consensus-driven business invariants. \tool systematically mines these invariants from large-scale on-chain contracts and reuses them as reference constraints to audit target contracts. To achieve this, \tool introduces a Business Specification Language (BSL) to normalize diverse implementation patterns into structured, verifiable logic representations. It further combines noise-aware logic aggregation with contrastive auditing to identify missing or weakly enforced invariants while mitigating LLM-induced false positives.

We evaluate \tool on three real-world datasets, including DeFiHacks, Web3Bugs, and a set of top-200 audited contracts. The results show that \tool achieves an F1 score of 85.2\%, significantly outperforming state-of-the-art tools while maintaining a low false-positive rate on production-grade contracts. Additional experiments demonstrate that \tool maintains consistent performance across different LLMs and is cost-effective, and that its false-positive suppression mechanisms substantially improve robustness. Overall, \tool provides a practical and scalable approach to reasoning about smart contract business logic through invariant coverage analysis.
\end{abstract}



\received{20 February 2007}
\received[revised]{12 March 2009}
\received[accepted]{5 June 2009}

\maketitle

\section{Introduction}
Smart contracts are automatically executable programs deployed on blockchain networks, enabling transactions to be completed automatically without intermediaries. They constitute the backbone of decentralized applications (dApps) and decentralized finance (DeFi). However, their immutability and public accessibility make security a critical concern. According to the 2025 Crypto Crime Security Report released by Chainalysis~\cite{chainalysis2025cryptoCrimeReport}, as of June 2025, cryptocurrency services have suffered security losses exceeding \$2.17 billion, surpassing the total losses of the entire year of 2024 and representing a 17\% increase compared to the same period in 2022. Further studies \cite{DBLP:journals/csur/LiuZWSZY26,ding2025comprehensivestudyexploitablepatterns} indicate that the attacks causing the greatest losses are not traditional vulnerabilities such as reentrancy \cite{DBLP:conf/ndss/Song0JXC25} or integer overflow \cite{chen2025numscout}, but rather business logic vulnerabilities. Such vulnerabilities typically manifest as missing critical condition checks or flawed design-level logic, and are strongly correlated with specific protocol semantics. As a result, they have become the most covert and hardest-to-detect type of vulnerability.

Existing smart contract vulnerability detection techniques can be roughly divided into static and dynamic analysis, and large language model (LLM)-based approach. Static analysis tools, such as Slither\cite{feist2019slither}, Mythril\cite{mythril2024}, and SmartCheck\cite{tikhomirov2018smartcheck} aims to identify fixed-pattern vulnerabilities by parsing abstract syntax trees (ASTs), control flow graphs (CFGs), data-flow analysis, or symbolic execution. Dynamic analysis tools, such as Echidna\cite{10.1145/3395363.3404366}, ContractFuzzer\cite{10.1145/3238147.3238177}, and SMARTIAN\cite{choi:ase:2021} observe runtime behaviors and potential risks by executing smart contracts. 
Recently, LLMs have been increasingly adopted in vulnerability detection for smart contracts. For instance, GPTScan\cite{sun2024gptscan}, PropertyGPT\cite{liu2024propertygpt}, and SmartInv\cite{wang2024smartinv} leverage the strong semantic understanding capabilities of LLMs to overcome the limitations of traditional tools in capturing business semantics.

Despite its promising results, existing tools face challenges in detecting logic vulnerabilities specific to high-level smart contract business. 
These vulnerabilities arise from protocol design flaws, e.g., missing invariant checks or incorrect state ordering, which lack universal detection patterns for static analyses to apply and rarely trigger exceptions detectable by dynamic analyses. 
Existing LLM-based approaches harness LLM to identify critical variable for pattern-based analysis~\cite{sun2024gptscan}, generate properties guided by expert-written specification ~\cite{liu2024propertygpt} and subtle arithmetic invariants ~\cite{wang2024smartinv}.
While achieving great success, they are struggling to generalize to unknown types of smart contracts without priori knowledge extracted from well-defined documentation \cite{ethereum_erc} or common development practice~\cite{openzeppelin_website}, thus failing to capture the complex business workflows required to systematically check business invariant enforcement.

To address the above challenges, in this paper, we proposes {\tool}, an automated framework for detecting business logic vulnerabilities in smart contracts through contrastive auditing. The core idea is that real-world on-chain projects contain rich and mature business logic. Instead of explicitly writing invariant rules, we automatically extract business logic templates from on-chain data and use them as reference invariants to examine critical business constraints in other contracts.
These invariants are consensus-based in nature and serve as security-relevant constraints for logic risk analysis.
Deviations from such consensus invariants are treated as review-worthy signals for potential vulnerabilities.

\tool consists of \emph{Logic Miner} and \emph{Logic Checker} for mining and checking smart contract business logic, respectively. {Logic Miner} systematically mines business logic from categorized on-chain contracts. It employs a lightweight LLM to distill code into structured Business Specification Language (BSL) representations, creating a database of verifiable consensus invariants. 
{Logic Checker} performs contrastive auditing. It retrieves semantically relevant reference implementations and associated invariants and guides an LLM to detect missing checks in the target code. Mechanisms like multi-dimensional code extraction and noise-aware aggregation are integrated to ensure robustness against implementation variations and minimize false positives.

We implement our approach and conduct an extensive evaluation on three high-quality real-world datasets: DeFiHacks, Web3Bugs, and the Top-200 contracts ranked by market capitalization. The experimental results show that \tool achieves an F1 score of 85.2\%, significantly outperforming existing state-of-the-art approaches. 
Particularly, it also suppresses hallucinations in LLM reasoning process, maintaining 7.1\% false-positive rate on production-grade contracts while detecting 94.3\% of vulnerable cases from DeFiHacks dataset. 
Our experiment results also confirm that \tool achieves consistent performance across different LLMs and remains cost-effective, making it a practical alternative to automating real-world smart contract auditing.

The main contributions of this paper are summarized as follows:
\begin{itemize}[leftmargin=*]
    \item We propose an LLM-driven framework that is able to detect business logic vulnerabilities for smart contracts by contrastive auditing, through inducing business logic invariants derived from real-world on-chain projects.
    \item We design a verifiable intermediate representation called Business Specification Language (BSL). Coupled with a dual validation mechanism, it enables the reliable extraction of structured logic from unstructured code, effectively mitigating LLM hallucinations.
    \item We implement our approach in \tool, developing a noise-aware logic aggregation method that overcomes code diversity and data noise to establish high-quality business logic knowledge base. We evaluate \tool on diverse datasets, achieving a high F1 score of 85.2\%, outperforming the state-of-the-art approaches.
\end{itemize}

The remainder of this paper is organized as follows. Section~\ref{sec:motivation} presents a motivating example to illustrate the problem. \Cref{sec:BSL} proposes the Business Specification Language (BSL). \Cref{sec:overview}  details the design of \tool. \Cref{sec:evaluation} presents the experimental evaluation. 
We discuss related work in \Cref{sec:related_work}, and conclude the paper in \Cref{sec:conclusion}.

\begin{figure}[h]
\centering
\begin{subfigure}[t]{0.45\linewidth}
\begin{lstlisting}[language=Solidity, basicstyle=\scriptsize\ttfamily, xleftmargin=2em, numbers=left,escapeinside={(*@}{@*)}]
function donateToReserves(..., uint amount) external {
    // 1. Basic Check: Balance
    uint bal = eToken.balanceOf(msg.sender);
    require(bal >= amount, "insufficient");

    // (*@\textcolor{red}{<--- VULNERABILITY:}@*)
    // (*@\textcolor{red}{Missing Health Factor Check}@*)

    // 2. Action: Burn user tokens
    eToken.burn(msg.sender, amount);
}
\end{lstlisting}
\caption{Vulnerable (Target Contract).}
\label{fig:euler}
\end{subfigure}
\hfill
\begin{subfigure}[t]{0.52\linewidth}
\begin{lstlisting}[language=Solidity, basicstyle=\scriptsize\ttfamily, xleftmargin=2em, numbers=left]
function executeWithdraw(..., uint amount) external {
    // 1. Basic Check: Balance
    uint bal = IAToken(aToken).balanceOf(msg.sender);
    require(bal >= amount, "insufficient");
    // 2.  Check Health Factor
    (uint debt, ) = getUserData(msg.sender);
    uint HF = calculateHF(debt, bal - amount);
    require(HF >= HF_THRESHOLD, "unsafe");
    // 3. Action: Burn user tokens
    IAToken(aToken).burn(msg.sender, amount);
}
\end{lstlisting}
\caption{Correct (Template Contract).}
\label{fig:poolInstance}
\end{subfigure}
\caption{Comparison of Business Logic: EulerFinance (missing check) vs. PoolInstance (enforcing invariant).}
\label{fig:motivation_example}
\end{figure}

\section{Motivation Example}
\label{sec:motivation}

To illustrate the impact of business logic vulnerabilities, we examine the Euler Finance incident from March 2023~\cite{euler-etoken-sol-fa93987}, which resulted in a staggering loss of approximately \$197 million.
\Cref{fig:euler} shows the vulnerable function of this contract. In the EToken contract, a \texttt{donateToReserves} function is provided, allowing the contract to burn or donate a portion of the user's tokens. Before burning or donating the tokens, the contract checks the user's balance to ensure that the balance is greater than the amount to be burned (line 4). It then subtracts the donation amount from the user's balance and increases the reserves (line 10). However, a core piece of logic missing in \texttt{donateToReserves} is that the contract ignores the user's debt or health factor when burning the user's tokens\footnote{The health factor reflects the relationship between a user's collateral and its debt; when the debt is too high relative to the collateral, the health factor decreases and may trigger liquidation.}. An attacker can actively burn their own tokens, reducing their collateral balance while their debt remains unchanged, which lowers the account's health factor and triggers the system's forced liquidation mechanism. The attacker then purchases the liquidated collateral at extremely low cost, obtains additional rewards, and generates large bad debt before exiting with profit.

In contrast, we examine the on-chain Pool Instance contract, which implements a correct reference for collateral reduction logic.
Although \Cref{fig:euler} (donation) and \Cref{fig:poolInstance} (withdrawal) differ in user intent, they share the same fundamental lending semantic: decreasing the user's collateral balance.
In a lending protocol, any action that reduces a user's collateral---whether withdrawing to a wallet, transferring to others, or donating to reserves---increases the user's leverage and risk of insolvency. Therefore, strictly enforcing the Health Factor check is a universal invariant for all such operations.
As shown in \Cref{fig:poolInstance}, the Pool Instance's \texttt{executeWithdraw} function enforces this check (line 10) before burning tokens, thereby preventing the vulnerability found in Euler.

Although the two examples differ in implementation details, they share the same business scenario: token burning in a lending protocol. Crucially, validating the user’s debt status (via Health Factor) before burning tokens constitutes a consensus business invariant implicitly encoded in mature protocols. Specifically, the vulnerable implementation omits this critical safeguard, whereas the reference example strictly enforces the invariant. Based on this observation, we propose the core idea of this paper: by inducing the business invariants that are commonly adopted within the same business scenario and detecting the absence of these invariants in the target contract, potential business logic vulnerabilities can be effectively identified.

\begin{figure}[h]
\small
\begin{align*}
	\emph{v} \in \; StateVar& \quad \emph{param} \in \; TemporalVar \\ \nonumber
    \emph{BusinessSC} &=  v*; BusinessFunc*; LogicFunc* \\ \nonumber 
    \emph{LogicFunc} &= stmt*\\ \nonumber
    \emph{BusinessFunc} &= checkCondition*; stmt*; \textbf{call}(LogicFunc); stmt* \\ \nonumber 
    \emph{\textbf{Spec}(BusinessFunc)} &= checkCondition*; \textbf{call}(LogicFunc) \nonumber \\
    \emph{stmt} &= checkCondition\; ||\; exp \;|| \; \textbf{call}(LogicFunc) \\ \nonumber
    \emph{checkCondition} &= \textbf{assert}(bool\_exp(v, param))\; || \textbf{require}(bool\_exp(v, param))\; \\ \nonumber &\quad||\; \textbf{if}\; (bool\_exp(v, param))\; \textbf{revert}() \nonumber
\end{align*}
\caption{Business Specification Language (BSL) for Solidity Smart Contracts.}
\label{fig:BSL}
\end{figure}

\section{Business Specification Language }\label{sec:BSL}

To generalize the observation in Section ~\ref{sec:motivation} into an automated auditing framework, we define the Business Specification Language (BSL), which captures the requirement that a protocol must enforce specific invariants before critical actions.
BSL serves as a normalized intermediate representation (IR) that bridges the gap between unstructured Solidity code and automated logic analysis.

As formally defined in Figure \ref{fig:BSL}, BSL provides a rigorous grammar to abstract smart contract business logic. At the top level, a \emph{BusinessSC} is composed of state variables ($v$), business entry functions ($BusinessFunc$), and internal logic implementations ($LogicFunc$). Crucially, BSL introduces a specification abstraction, denoted as $Spec(BusinessFunc)$, which distills the complex control flow of a function into two core components:

\begin{itemize}[leftmargin=*,topsep=0pt]
    \item Precondition Constraints ($checkCondition*$): This component normalizes diverse validation patterns found in Solidity source code. As defined in the last rule of Figure~\ref{fig:BSL}, BSL standardizes all guard clauses into a unified \texttt{checkCondition} format. This applies to various implementations, including assert, require, and if...revert. To ensure consistency, BSL specifically inverts negative logic (e.g., if...revert) into positive assertions.
    
    \item Semantic Action ($call(LogicFunc)$): This represents the core business operation (e.g. token transfer, state update) that executes only after all preconditions are satisfied.
\end{itemize}

This design offers distinct advantages for automated auditing. First, normalizing diverse patterns into positive preconditions decouples business invariants from implementation details. This ensures consistent representation of functionally equivalent checks, enabling accurate cross-protocol matching. Second, BSL’s lightweight constrained grammar minimizes ambiguity compared to complex formal languages, ensuring deterministic logic extraction that is directly applicable to efficient downstream invariant analysis. In particular, we design BSL to focus on high-level business invariants rather than low-level arithmetic operations.

\section{LogicScan}\label{sec:overview}

\begin{figure}[h]
  \centering
  \includegraphics[width=\linewidth]{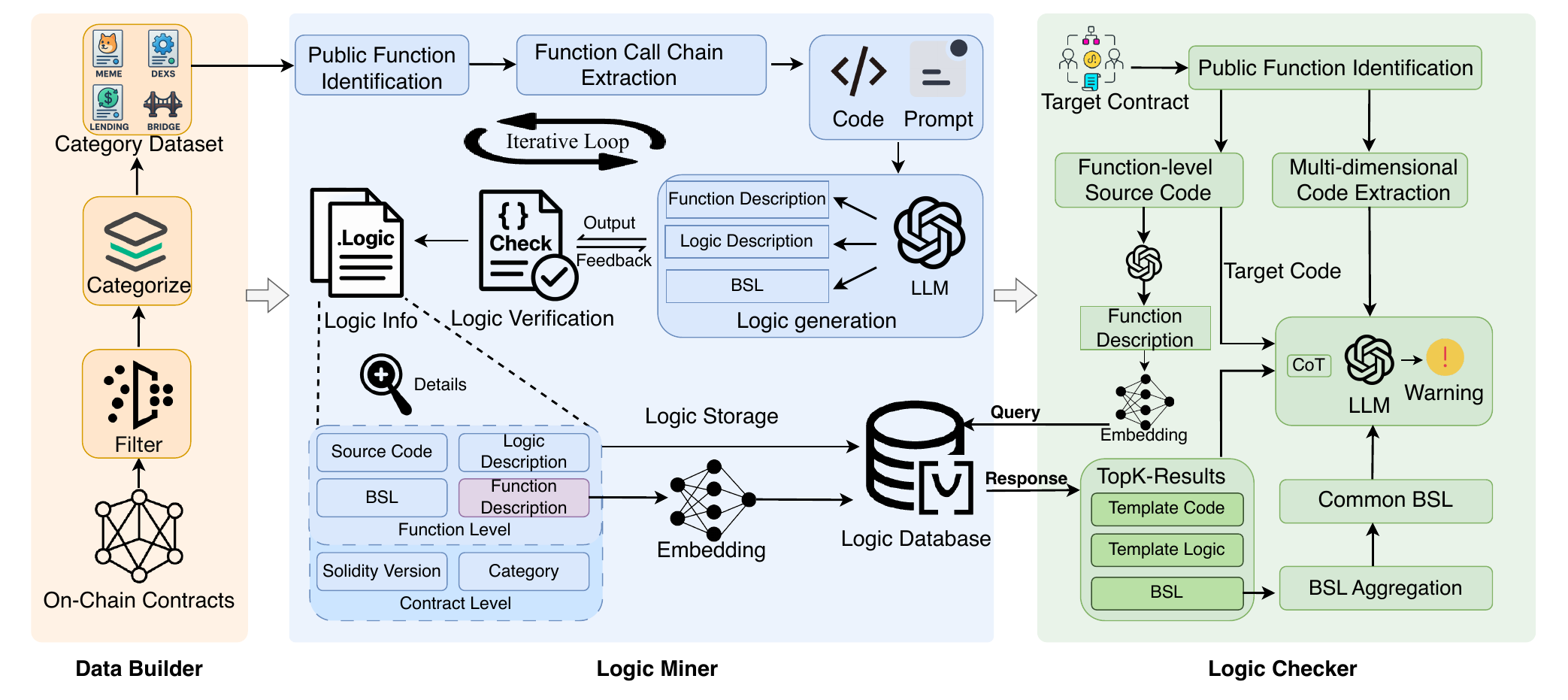}
  \caption{A workflow of LogicScan.}
  \label{fig:overview}
\end{figure}

\subsection{Overview}
\tool is an automated framework for detecting business-logic vulnerabilities in smart contracts by detecting the omission of critical business invariants.
Its key insight is that widely deployed on-chain protocols implicitly encode well-tested business logic constraints, which can be mined and reused as reference invariants for auditing other contracts.
As illustrated in Figure~\ref{fig:overview}, \tool consists of two core components: {Logic Miner} and {Logic Checker}.

\emph{Logic Miner} extracts reusable business-logic invariants from real-world on-chain contracts.
Given a large corpus of category-labeled contracts, Logic Miner analyzes publicly accessible functions and distills function-level business semantics.
A lightweight LLM is used to generate concise functional and logical descriptions, which are then normalized into a structured representation using a Business Specification Language (BSL).
Logic Miner performs both syntactic and semantic validation on the extracted BSL to reduce the randomness and hallucination of LLM outputs. Based on the validated BSL, we construct a domain knowledge dictionary, which is later used to capture common logic patterns across similar function implementations. Finally, we design a structured storage schema to support more accurate and efficient retrieval.

\emph{Logic Checker} performs contrastive auditing on target contracts using the mined invariants.
For each externally callable function in the target contract, Logic Checker first extracts its call-chain code and generates a functional description.
It retrieves semantically similar logic templates from the business logic database and guides the LLM to perform contrastive auditing by comparing the target implementation with template implementations in the same scenario.
To reduce false positive rates, we introduce Multi-dimensional Code Extraction and Noise-aware Logic Aggregation across Templates, thereby enhancing the LLM's ability to understand specific scenarios.

\subsection{On-chain Data Collection for Smart Contracts}\label{sec:sc_collection}

\tool relies on the observation that mature on-chain protocols implicitly encode well-tested business invariants through their implementation logic. To mine such invariants, we first construct a large-scale corpus of real-world smart contracts that are both actively used and representative of established business scenarios.
The process of the collection is illustrated in Part 1 of Figure~\ref{fig:overview}.

\textbf{Contract Filtering.} We collect verified smart contracts deployed on Ethereum and apply a series of filters to retain contracts that are more likely to reflect stable and meaningful business logic.
Specifically, we remove contracts that are inactive, proxy-only, interfaces, or libraries, as these either do not contain executable business logic or merely forward calls to other implementations.
We further prioritize contracts that have been deployed for a sufficiently long time and exhibit recent on-chain activity, under the assumption that frequently used and long-lived contracts tend to undergo more extensive testing and real-world validation.
Finally, we kept top 20\% contracts both in deployed time and transaction volume, which are 43,749 out of 817,772 contracts.
This filtering step yields active contracts that serve as reliable sources for inducing business-logic constraints.

\begin{figure}[h]
  \centering
  \includegraphics[width=\linewidth]{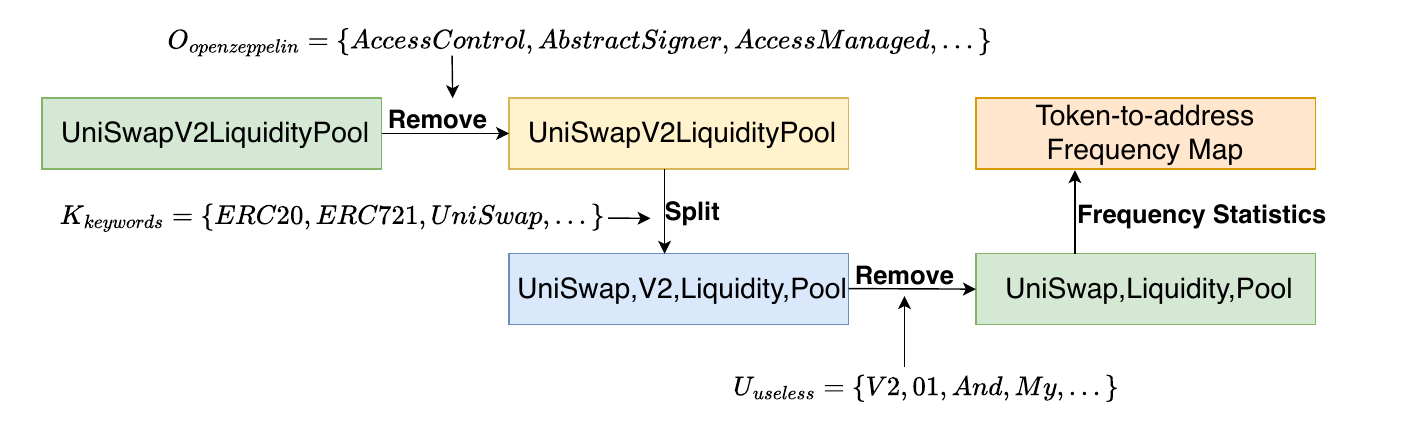}
  \caption{Contract Categorization Process.}
  \label{fig:Contract-Categorization-Process}
\end{figure}

\textbf{Contract Categorization.}
Since business invariants are inherently scenario-dependent, contracts from different application domains cannot be directly compared.
We therefore assign coarse-grained business category labels to the filtered contracts using a static, name-based classification strategy.
In practice, smart contract names often encode strong semantic signals about their intended business roles (e.g., \texttt{Token}, \texttt{Pool}, \texttt{Vault}, \texttt{Farm}), following camel case and snake case naming conventions commonly used in Solidity.
We tokenize contract names, remove trivial or version-related terms (defined in a predefined stop-word list), and retain domain-specific keywords that correlate with business functionality.
Figure~\ref{fig:Contract-Categorization-Process} illustrates the categorization workflow using a representative contract name as an example.
Based on the resulting token-to-address map, contracts associated with the same high-frequency category tokens are grouped into the same business category.

This categorization approach is intentionally lightweight, aiming to ensure coarse scenario alignment rather than perfect classification. Importantly, \tool does not rely on any single category decision: multiple template functions are retrieved and their invariants are aggregated through frequency-based voting. As a result, occasional misclassifications only introduce noise but are unlikely to dominate the final consensus invariants unless they are systematic across many templates.
This labeled corpus forms the foundation for Logic Miner to induce statistically representative business-logic invariants from real-world deployments.

\subsection{Logic Mining Pipeline: Generation, Verification, and Repair}\label{sec:logic-Mining-Pipeline}

To instantiate the BSL templates defined in Section~\ref{sec:BSL} and mitigate LLM hallucinations, Logic Miner employs a Generate-Validate-Repair pipeline.
A lightweight LLM generates the BSL specification based on the function's source code. The strictly constrained BSL grammar significantly reduces the model's output space, lowering randomness.

Logic Miner is designed to generate invariants from popular on-chain projects for detecting logic vulnerabilities in Logic Checker.
Logic Miner operates in two stages.
First, it collects popular on-chain projects and assigns them to coarse-grained business categories.
Secondly, for each business category, a business logic template is generated to detect potential logic flaws.

Given a category-labeled corpus of on-chain smart contracts, Logic Miner extracts reusable
business-logic knowledge through a unified logic mining pipeline. The pipeline transforms raw
contract code into explicit, normalized, and verifiable business invariants, enabling robust
downstream analysis and comparison across heterogeneous implementations.

\begin{figure}
\centering
\label{logic_miner_prompt}

\begin{tcolorbox}[title=Logic Miner Prompt, boxsep=1pt, left=2pt, right=2pt, top=2pt, bottom=2pt]
\small
\textbf{R1 (Func. Desc.):} Analyze function \texttt{\{name\}} and call chain \texttt{\{code\}} in a DeFi \texttt{\{category\}} contract. Generate a 1-sentence description following: ``This function is used to \textit{<purpose>}, accepts \textit{<inputs>}, processes \textit{<logic>}, and outputs \textit{<result>}.''

\tikz{\draw[dashed, gray, line width=0.5pt] (0,0) -- (\linewidth,0);}

\textbf{R2 (Logic Summary):} Summarize mandatory pre-execution checks (focusing on \texttt{require}, \texttt{assert}, \texttt{revert}) into a single sentence. Output only the summary without explanation.

\tikz{\draw[dashed, gray, line width=0.5pt] (0,0) -- (\linewidth,0);}

\textbf{[R3] BSL Generation:} Convert checks to BSL (syntax: \texttt{\{BSL\}}). Format: \texttt{order(check[cond1, ...], action)}. \textbf{Rules:} 1) Map explicit validation to positive preconditions \textbf{(invert \texttt{if...revert} logic)}. 2) Action reflects semantic purpose. \textbf{Example:} \texttt{order(check[balance\_ge\_reserve], swap)}.

\tikz{\draw[dashed, gray, line width=0.5pt] (0,0) -- (\linewidth,0);}
\textbf{[R4] Logic Repair:} The generated BSL failed validation. \textbf{Error:} syntax error at \texttt{\{error\_message\}}; count mismatch: source \texttt{\{actual\_count\}} vs BSL \texttt{\{bsl\_count\}}.
\textbf{Task:} Correct the \texttt{\{invalid\_bsl\}} based on the error. Ensure strict adherence to BSL syntax and that the number of checks matches the source code exactly. Output only the corrected BSL.
\end{tcolorbox}
\vspace{-1em} 
\caption{Logic Miner Prompt.}
\label{fig:logic_miner_prompt}
\end{figure}

\textbf{Generation.}
For each publicly accessible function, Logic Miner first statically extracts the complete internal call-chain implementation by expanding all transitively invoked internal functions. It then characterizes the function from six complementary aspects, each capturing a distinct dimension of its business logic:
\begin{enumerate}[leftmargin=*]
    \item Complete internal call-chain implementation: captures the full executable behavior of the function by preserving all transitively invoked internal calls, ensuring that the entire execution context is analyzed rather than a single entry point.
    \item Functional description: summarizes the intended purpose and high-level behavior of the function at the semantic level, abstracting away implementation details.
    \item Logical description: emphasizes the explicit validation logic and invariant checks enforced during execution, such as preconditions and guard conditions.
    \item BSL grammar: provides a normalized and structured abstraction of the logical description using the BSL, facilitating subsequent static analysis and invariant comparison.
    \item Business category label: situates the function within its financial role and application scenario, enabling logic aggregation and comparison within the same business scenario.
    \item Solidity version information: solidity version information reflects inherent differences in security design, and taking these design variations into account helps reduce false positives.
\end{enumerate}

Among these aspects, (1), (5), and (6) are obtained through static analysis. Aspects (2), (3), and (4) require semantic understanding of source code and are therefore generated using a lightweight LLM. The corresponding three rounds of prompts are shown in ~\ref{fig:logic_miner_prompt}.

Crucially, Logic Miner restricts logic extraction to explicitly stated validation logic, including \texttt{require statements}, \texttt{assert}, and \texttt{revert-guarded branches}. The model is strictly instructed not to infer implicit checks or unstated assumptions. All extracted logic must be directly traceable to concrete source-code constructs. This design choice mitigates hallucinated conditions and enables subsequent verification. 

\textbf{Verification.}
To ensure the correctness of the generated logic, we enforce a two-stage validation mechanism. For syntactic validation, we implement a strict parser (e.g., based on ANTLR4) to check if the LLM output conforms to the BSL grammar. Any parsing error triggers an immediate failure with specific location feedback.
For semantic validation, to filter out hallucinations, we verify the consistency between the generated specification and the source code. Specifically, we extract explicit validation statements (e.g., \texttt{require}, \texttt{assert}, \texttt{if-revert}) from the AST and compare their count against the \texttt{checkCondition} entries in the BSL. Discrepancies in quantity trigger an error report containing the actual count found in the source.

\textbf{Repair.} 
Error messages from both validators are fed back to the LLM(see R4 in Figure~\ref{fig:logic_miner_prompt}) for iterative refinement. To prevent infinite loops, we limit this process to four iterations.If the BSL remains invalid after these attempts, LogicScan simply skips the function to proceed with the remaining analysis. This closed-loop process effectively transforms probabilistic LLM outputs into deterministic and verifiable logic representations.

Although LLMs generate BSL representations that satisfy syntactic and basic semantic correctness, their semantic expressions still exhibit substantial variability (e.g., \texttt{caller\_is\_owner}, \texttt{only\_owner}, \texttt{owner\_check}, \texttt{isOwner}). To address this issue, we design and construct a domain knowledge dictionary that canonicalizes logically equivalent attributes into unified representations, eliminating superficial lexical differences.
Specifically, we extract all logical check conditions appearing in the BSL corpus, yielding more than 62{,}000 distinct invariant conditions. We then compute their occurrence frequencies and select the top 2,000 most frequent invariant conditions, which empirically provides a good balance between coverage and noise reduction. Based on semantic similarity among these top conditions, we construct a domain knowledge–driven synonym dictionary through lightweight manual curation. This dictionary prevents semantically equivalent invariants from being fragmented into separate representations and substantially improves the stability of common logic aggregation.

To support accurate and efficient retrieval of function-level logic information, we design a hybrid storage architecture that combines a vector database with a relational database. Specifically, we adopt an open-source embedding model to vectorize the (2) functional descriptions of functions and store these embeddings in a vector database for similarity-based retrieval. The remaining logic features are stored in a relational database, and the two storage systems are linked via unique function identifiers.

Therefore, this unified pipeline allows Logic Miner to distill noisy, heterogeneous contract implementations into stable and reusable business-logic templates and store them in a logical database, which serves as the foundation for contrastive auditing in Logic Checker.

\subsection{Retrieval of Business Logic Invariant}

To support business-logic risk analysis for a target contract, Logic Checker requires function-level \emph{template functions} that implement the same functionality as the target functions. These template functions serve as reference knowledge for the LLM during auditing, enabling Logic Checker to compare the target implementation against consensus business invariants learned from real-world deployments and to produce interpretable analysis reports.

Existing function retrieval methods for smart contracts typically rely on name-based matching or embedding similarity over source code or AST representations (e.g., ZepScope~\cite{liu2024using}). These approaches are highly sensitive to naming conventions and syntactic structures, and often fail to retrieve semantically equivalent functions implemented using different coding patterns or identifiers, resulting in limited robustness and precision.

To address the limitations of name-based and code-structure–based retrieval, we propose a function retrieval method based on semantic similarity of high-level functionality. As illustrated in the third part of Figure~\ref{fig:overview}, given a target contract, we retrieve reference functions by jointly considering three complementary dimensions: the Solidity version of the contract, the complete call-chain implementation of each publicly callable function, and the contract’s business category.
This design is motivated by three considerations. First, Solidity version information is incorporated to avoid mismatches caused by cross-version differences in language semantics and built-in security mechanisms. Second, semantic retrieval based on complete internal call chains enables accurate matching even when equivalent functionality is implemented using different coding patterns or naming conventions. Third, retrieval is restricted to the same business category.
This avoids comparing functions that share similar interfaces but implement inherently different domain-specific semantics.
In practice, given a target contract, Logic Checker extracts its Solidity version, determines its business category, and identifies all publicly callable functions. For each function, it expands the complete internal call chain and generates a concise functional description using the same LLM pipeline as Logic Miner.
The resulting descriptions are vectorized using the same embedding model as in Logic Storage, and semantic retrieval is performed via cosine similarity against the logic database. Retrieved candidates are ranked by similarity and filtered to retain only functions within the same Solidity major version and business category.

Due to semantic variability in LLM-generated functional descriptions, a similarity threshold is required for stable retrieval. We conduct an empirical stability analysis.
We observe that correct self-retrieval occurs within the top five results in over 90\% of cases.
Based on this observation, Logic Checker adopts top-5 as the retrieval threshold. In addition, candidate templates with a similarity score below 80\% are discarded to avoid unreliable matches.

\subsection{Business Logic Consistency Checking}\label{sec:business_logic_vul_detect}

The goal of consistency checking is to verify whether a target function explicitly enforces the essential business constraints commonly adopted by semantically equivalent implementations. Unlike traditional vulnerability detection that focuses on low-level coding errors, this analysis compares the target implementation against consensus invariants derived from real-world deployments. Importantly, missing invariants are treated as review-worthy signals rather than definitive vulnerabilities, as deviations may also reflect benign design choices.

This consistency checking process is performed under the assumption that a set of representative reference implementations for the target function is already available. Specifically, for a given target function, Logic Checker retrieves the top-$k$ template functions that are \emph{likely} to be semantically equivalent in functionality, constrained by Solidity version compatibility and business category consistency. Each template function is associated with its source code, calling context, and extracted business-logic representations, which together capture consensus business invariants observed across real-world deployments.

These retrieved template functions serve as reference knowledge that reflects how similar functionality is typically implemented in practice. 

\textbf{Consistency Checking Procedure.}
Given the reference template functions and the target implementation, Logic Checker performs this consistency verification process through a structured, three-phase procedure. This procedure is operationalized using a LLM through controlled, multi-round interactions, where each phase incrementally guides the model to reason over reference implementations and the target code under explicit constraints.

\begin{itemize}[leftmargin=*,topsep=0pt]
\item \emph{Invariant Induction.} Logic Checker summarizes common business invariants from retrieved template code, capturing consensus constraints grounded in real-world implementations. This process establishes a security benchmark within the business category.

\item \emph{Consistency Verification.} The target function is then analyzed to determine whether each invariant is explicitly enforced or semantically preserved, allowing logically equivalent checks implemented in different coding styles.

\item \emph{Deviation Analysis.} Finally, uncovered or partially satisfied invariants are interpreted to distinguish benign design deviations from omissions that may introduce exploitable or inconsistent behaviors.

\end{itemize}

Across these phases, the LLM is not used to infer new business rules, but to assist in summarizing, aligning, and reasoning about explicitly observed invariants grounded in real-world implementations.

\subsection{Mitigation Strategies of LLM's Hallucination }\label{sec:llm_result_opt}

After the business logic matching in the previous stage, false positives remain, primarily due to two reasons. First, a major source of these false alarms is incomplete contextual information, where some validation logic resides in functions related to variable assignments. Therefore, \tool incorporates a Multi-dimensional Code Extraction mechanism into its overall architecture to systematically mitigate these false positives.
The second is due to the existence of conflicted rules in the knowledge database, caused by redundant security designs.
A majority voting based rule aggregation is performed to mitigate the false positives.
This section provides a detailed explanation of the design motivations and their corresponding mechanisms.

\textbf{Context Completion via Multi-dimensional Code Extraction.}
To prevent LLMs from producing incorrect judgments in permission and state analysis due to missing context, we introduce a multi-dimensional code extraction mechanism in our design.
\tool not only obtains the complete call-chain code of the function but also further identifies variables involved in the call chain. Specifically, \tool identifies all variables contained in the call-chain code and retrieves all other functions that involve write operations to these variables. 
After enabling the LLM to read the function code related to variable write operations, this false-positive issue is significantly mitigated, enhancing the overall accuracy of the detection.

\newcommand{\ttwrap}[1]{\texttt{\seqsplit{#1}}}

\textbf{Noise-aware Logic Aggregation across Templates.}
Although the top-$k$ retrieved template functions share similar business scenarios and core functionalities, implementation patterns occasionally vary across developers, introducing redundant or conflicting logic branches. If a LLM treats these minority, implementation-specific variations as mandatory constraints during contrastive auditing, the false-positive rate increases significantly. To address this, we propose a voting-based business logic aggregation algorithm designed to eliminate implementation-induced noise and reinforce the model’s understanding of \textbf{common invariants}. The methodology consists of three stages:

\begin{itemize}[leftmargin=*,topsep=0pt]
\item \textbf{checkCondition Extraction}: Utilizing a custom-developed BSL grammar parser, we extract each pre-execution guard condition (\texttt{checkCondition}) from the retrieved BSL specifications.

\item \textbf{Semantic Normalization}: Based on the domain-driven synonym dictionary detailed in Section~\ref{sec:logic-Mining-Pipeline}, we perform synonym substitution—for example, mapping \texttt{is\_owner} and \texttt{only\_owner} to a canonical term—to eliminate lexical variances across diverse implementation patterns.

\item \textbf{Redundancy and Conflict Filtering}: We apply a consensus-based voting mechanism to filter normalized logic and identify the most representative business specifications. Specifically, the occurrence frequency of each invariant is calculated across the top-$k$ set, and invariant that fail to secure majority support (i.e., frequency $< \lceil k/2 \rceil$) are deemed project-specific outliers or redundant logic and are subsequently pruned. For invariants possessing antonymous attributes defined in the domain knowledge dictionary (e.g., \texttt{\_not\_excluded} vs.\ \texttt{\_excluded}), the system evaluates their distribution: if the counts are balanced (indicating a lack of clear consensus), the conflicting pair is removed to avoid ambiguity; otherwise, the majority rule is retained while the minority conflict is discarded.
\end{itemize}

Through this process, we eliminate redundant logic to derive a common business logic template. Subsequently, we integrate the complete code and this common template into the prompt design. This reinforces and calibrates the LLM's understanding of common business invariants, enabling it to rectify previously generated outputs.

\section{Evaluation}\label{sec:evaluation}
In this section, we evaluate \tool on diverse datasets across different LLMs to answer the following research questions:
\begin{itemize}[leftmargin=*]
    \item \textbf{RQ1:} \textbf{(Effectiveness)} How effectively does \tool detect logic vulnerabilities for real-world smart contracts?
    \item \textbf{RQ2:} \textbf{(Sensitivity Study)} How effectively does \tool perform across different LLMs?
    \item \textbf{RQ3:} \textbf{(Efficiency)} What are the runtime performance and cost overheads of \tool?
    \item \textbf{RQ4:} \textbf{(Ablation Study)} How effectively do \tool's core components reduce false positives?
    
\end{itemize}

\textbf{Datasets.} 
We collected three datasets to evaluate \tool.
To investigate the effectiveness of \tool in terms of vulnerability detection, we curated two datasets from DeFiHacks\footnote{\url{https://github.com/SunWeb3Sec/DeFiHackLabs}} and Web3Bugs~\cite{DBLP:conf/icse/ZhangZXL23}.
The publicly available \textit{DeFiHacks} dataset contains real-world smart contracts that have suffered attacks. 
Among them, we manually selected contracts of which the labeled root causes are related to business logic flaws, resulting in a total of 52 vulnerable contracts containing 70 function-level vulnerabilities.
The dataset from \textit{Web3Bugs}~\cite{DBLP:conf/icse/ZhangZXL23} aggregates vulnerable contract projects with vulnerability audit reports published on Code4rena\cite{Code4rena}. 
Among the 81 contract projects initially identified in the Web3Bugs dataset, 16 projects were excluded as they could not be successfully parsed by crytic-compiler. Consequently, our final evaluation was conducted on 65 projects, which contained a total of 134 function-level vulnerabilities.
To evaluate false positives of \tool, we gathered a refined contract dataset from GPTScan~\cite{sun2024gptscan}, i.e., \textit{Top200}, which comprises the top 200 smart contracts ranked by market capitalization, excluding proxy smart contracts. 
These contracts are mostly audited and run for a long period. 
Their functionality has been explored in millions of transactions. 
Therefore, we treat them as benign smart contracts free from logic vulnerabilities.

\textbf{Baselines.} 
To the best of our knowledge, a few attempts have been made to detect logic vulnerabilities in smart contracts with LLMs, such as GPTScan~\cite{sun2024gptscan} and PropertyGPT~\cite{liu2024propertygpt}.
However, since PropertyGPT does not have publicly available artifacts, we only compare \tool with {GPTScan}, which is the current SOTA solution that combines LLMs with static analysis tools.
To ensure a fair comparison and minimize experimental bias, we configured GPTScan using the same Large Language Model (GPT-3.5-turbo~\cite{openai_gpt35_turbo_docs}) and hyperparameters as specified in its original paper.
Additionally, we compare \tool with traditional vulnerability detection tools, including {Slither}~\cite{feist2019slither} that is an widely-used static analysis tool, and \textbf{ZepScope}~\cite{liu2024using} that statically detects target code by leveraging known design patterns in OpenZeppelin~\cite{openzeppelin_website} libraries. 
To study the gains of \tool beyond LLMs, we also compare \tool with one of the advanced LLMs to date, i.e.,{GPT-5.2 thinking} on the detection of logic vulnerability of smart contracts, aiming to confirm the validity of our proposed contrastive-based auditing approach.

To ensure comparability across tools with different reporting granularities, all results are normalized to the function level. For baseline tools that report issues at the contract level, we implemented a location-based mapping strategy: a reported issue was mapped to specific functions by tracing the source code coordinates.
In our evaluation, we use \emph{precision}, \emph{recall}, and \emph{the F1 score} as evaluation metrics to compare \tool with the aforementioned detection tools.

\subsection{Experiment Setup}
We conducted our experiments on a workstation equipped with an AMD Ryzen 9 5900X CPU, 64GB RAM, and an NVIDIA RTX 4080S (16GB) GPU. To ensure reproducible and efficient analysis, we employed a tiered inference strategy that distributes tasks across models of different scales.

For the embedding and retrieval process, we utilized Crytic-compile (v0.3.10) for smart contract compilation. The E5-Mistral-7B-Instruct model\cite{wang2024improvingtextembeddingslarge} was employed to generate semantic embeddings for function retrieval, which were subsequently stored in a FAISS vector database.

To implement the tiered reasoning strategy and balance performance with cost, we offloaded low-level semantic extraction tasks—including the generation of one-sentence functional descriptions, logical descriptions, and BSL—to a locally hosted gpt-oss-20b model~\cite{openai_gptoss20b}. In contrast, the more complex contrastive auditing and high-level logic reasoning were conducted via OpenRouter APIs\cite{OpenRouter} using advanced models such as GPT-5.

\subsection{RQ1: Effectiveness in vulnerability detection}

Vulnerability detection tools, both static and LLM-based, usually suffer from high false positive rates and limited recall~\cite{sun2025llm4vulnunifiedevaluationframework,li2024static}, 
making them impractical to be applied to real-world scenarios. 
In RQ1, our objective is to compare the precision and false positive rate of \tool compared with other state-of-the-art (SOTA) tools on the DeFiHacks, Web3Bugs, and Top200 datasets. 

\Cref{tab:rq1_results} shows the experiment results.
The first two columns represent the various evaluation datasets (DeFiHacks, Top200, and Web3Bugs) and the auditing tools used for comparison. The rest columns display several key statistical metrics: TP, FP, and FN record the specific counts of True Positives, False Positives, and False Negatives, respectively. Furthermore, Precision, Recall, and F1-score provide standardized performance percentages to facilitate a direct comparison of the strengths and weaknesses of \tool against GPTScan, ZepScope, Slither, and the LLM-Only baseline.
On the DeFiHacks dataset, \tool achieves the highest recall, detecting 66 out of 70 vulnerable functions and missing only 4 cases. It also reports substantially fewer false positives than the LLM-only baseline (19 vs. 192). As a result, \tool attains the best overall precision, recall, and F1 score (77.6\%, 94.3\%, and 85.2\%, respectively), outperforming all other approaches.
On the Top200 dataset, which consists of widely used and extensively audited contracts, all tools report zero true positives. 
GPTScan yields the fewest false positives with 12, while \tool reports a comparable figure of 18, and both are substantially lower than the false positive counts of other baselines. Although GPTScan reports the fewest false alarms across the datasets, \tool maintains a similarly low false positive rate on production-grade contracts while achieving significantly higher recall.
On the Web3Bugs dataset, \tool again achieves the best detection performance, identifying 111 vulnerable functions with only 23 misses. It outperforms all baselines in terms of precision (66.5\%), recall (82.8\%), and F1 score (73.7\%), demonstrating consistent effectiveness across different sources of real-world logic vulnerabilities.
Overall, across all three datasets, \tool achieves the best balance between precision and recall, indicating that it can preserve strong detection capability without a significant increase in false positives.

Although  \tool achieves satisfactory accuracy, investigating the root causes of false positives is critical for a comprehensive assessment. Analyzing the 75 false positives reveals that design-pattern deviation (38/75) was the primary cause. Beyond this, we identified distributed invariant enforcement (13/75), where validation is offloaded to upstream callers—for example, a transfer function omitting a non-zero address check because it was already enforced at the entry-level proxy function. We also identified implicitly encoded invariants (7/75) that rely on language semantics rather than explicit checks, such as relying on Solidity’s native arithmetic underflow protection to revert a transaction when a withdrawal exceeds a balance, instead of using an explicit require statement. The remaining 17 cases stemmed from other factors. Among these, we discuss a representative design-pattern deviation case.

\begin{figure}[h]
  \centering
  \begin{minipage}{\linewidth}
    \begin{lstlisting}[language=Solidity, xleftmargin=2.5em]
// Majority pattern: explicitly forbid self-approval
require(operator != msg.sender, "approve to caller");
// Target design: allow self-approval, only reject invalid operator
require(operator != address(0), "invalid operator");
    \end{lstlisting}
  \end{minipage}
  \caption{Case 1: Deviation from consensus-derived invariants (detected as False Positive).}
  \label{fig:case1-design-mismatch}
\end{figure}

\textbf{A false positive due to design-pattern deviation.} \Cref{fig:case1-design-mismatch} demonstrates how false positives reported by \tool can reveal meaningful deviations from historically dominant design patterns.
In the NFT protocol category, the statistical consensus mined by \tool shows that most deployed ERC721 contracts explicitly prohibit \textit{self-approval} in the \texttt{setApprovalForAll} function.
This pattern is consistently observed across several widely used projects, including \textit{PerilousPetz}\cite{etherscan:PerilousPetz:0x2af3}, \textit{BoringApes}\cite{etherscan:BoringApes:0xb29e}, \textit{GoblinRocks}\cite{etherscan:GoblinRocks:0xf9cf}, and \textit{DarkZodiacCollectible}\cite{etherscan:DarkZodiacCollectible:0xcb52}, which all enforce a guard ensuring that the operator address differs from the token owner, as illustrated in Figure~\ref{fig:case1-design-mismatch}.

The target contract\cite{etherscan:0x8877:0x8877}, however, adopts an alternative yet standard-compliant design choice.
Instead of forbidding self-approval, it permits this behavior and only enforces a minimal safety constraint that the operator is not the zero address.
Although this implementation deviates from historically prevalent practices, it preserves correct authorization semantics and does not introduce exploitable behaviors under the ERC721 specification.
Nevertheless, \tool flags this contract due to its violation of the consensus-derived “No Self-Approval” invariant.

From a security standpoint, this report constitutes a false positive.
However, rather than indicating a limitation of the approach, this case highlights \tool’s ability to surface departures from widely adopted community practices.
Such deviations may correspond to intentional design evolution, as also reflected in the transition from earlier to more recent versions of reference implementations such as OpenZeppelin’s ERC721.
Although such cases are counted as false positives in our evaluation, they represent review-worthy deviations from consensus designs and thus provide actionable signals to auditors for identifying non-standard or evolving contract logic.

\emph{Limitation.}
\tool focuses on pattern matching and local semantic analysis within a single function call, making it effective at detecting logic flaws such as missing access control or incorrect state-update ordering. However, \tool fails to detect subtle arithmetic bugs in smart contract. 
For instance, Balancer v2~\cite{OpenZeppelinBalancerV2Exploit2025}--style precision bugs are inherently \emph{sequence-amplified}. The rounding error introduced in any single computation is typically negligible and does not violate common templates or local invariants. An attacker can nonetheless exploit batched or multi-step interactions (e.g., \texttt{batchSwap}) to accumulate these small biases over many operations, eventually yielding exploitable profit. Detecting such vulnerabilities requires reasoning across multi-step execution traces and tracking accumulated rounding effects, which is beyond the scope of our current static, local-semantics--oriented analysis, leading to false negatives for this class of issues.

\begin{table}[t]
\centering
\small
\caption{Function-level evaluation results on DeFiHacks, Top200, and Web3Bugs.}
\label{tab:rq1_results}
\begin{tabular}{c l c c c c c c}
\toprule
Dataset & Tool & TP & FP & FN & Precision (\%) & Recall (\%) & F1-score (\%) \\
\midrule
\multirow{5}{*}{DeFiHacks}
& LogicScan & \textbf{66} & 19 & \textbf{4} & \textbf{77.6} & \textbf{94.3} & \textbf{85.2} \\
& GPTScan   & 16 & \textbf{14} & 44 & 53.3 & 26.7 & 35.6 \\
& ZepScope  & 4  & 217 & 66 & 1.8  & 5.7  & 2.7 \\
& Slither   & 1  & 285 & 69 & 0.3  & 1.4  & 0.5 \\
& LLM-Only  & 43 & 192 & 27 & 18.3 & 61.4 & 28.2 \\
\midrule
\multirow{5}{*}{Top200}
& LogicScan & 0 & 18 & 0 & -- & -- & -- \\
& GPTScan   & 0 & \textbf{12} & 0 & -- & -- & -- \\
& ZepScope  & 0 & 48 & 0 & -- & -- & -- \\
& Slither   & 0 & 53 & 0 & -- & -- & -- \\
& LLM-Only  & 0 & 40 & 0 & -- & -- & -- \\
\midrule
\multirow{5}{*}{Web3Bugs}
& LogicScan & \textbf{111} & 56 & \textbf{23} & \textbf{66.5} & \textbf{82.8} & \textbf{73.7} \\
& GPTScan   & 24  & \textbf{18} & 110 & 57.1 & 17.9 & 27.3 \\
& ZepScope  & 1   & 263 & 133 & 0.4 & 0.7 & 0.5 \\
& Slither   & 0   & 341 & 134 & 0.0 & 0.0 & 0.0 \\
& LLM-Only  & 52  & 231 & 64  & 18.4 & 44.8 & 26.1 \\
\midrule
\multirow{5}{*}{\textbf{Overall}}
& \textbf{LogicScan} & \textbf{177} & 75  & \textbf{27}  & \textbf{70.2} & \textbf{86.8} & \textbf{77.7} \\
& GPTScan   & 40 & \textbf{32}  & 154 & 55.6 & 20.6 & 30.1 \\
& ZepScope  & 5  & 480 & 199 & 1.0  & 2.5  & 1.4 \\
& Slither   & 1  & 626 & 203 & 0.2  & 0.5  & 0.2 \\
& LLM-Only  & 95 & 423 & 91  & 18.3 & 51.1 & 26.9 \\
\bottomrule
\end{tabular}
\end{table}

\answer{1}{
Across all three datasets, \tool consistently achieves higher precision and recall than existing tools, demonstrating its effectiveness in detecting business logic vulnerabilities while controlling false positives. False positives primarily arise from semantically valid design deviations that differ from common templates, reflecting \tool’s conservative auditing strategy. False negatives mainly correspond to sequence-amplified vulnerabilities requiring multi-step reasoning, which are beyond the scope of the current single-call–oriented analysis.
}

\subsection{RQ2: Effectiveness across various LLMs}

In RQ2, we aim to evaluate the model-agnosticism of this method. Specifically, we investigate whether \tool depends on the reasoning capability of a specific LLM, and whether it can be effectively adapted to open-source models.
We replace the default backend with three representative models: GPT-5 (closed-source/high-performance), Claude-sonnet 4.5 (long-context advantage), and Qwen-3 235B (open-source). We repeat the experiments on the aforementioned DeFiHacks and Web3Bugs datasets.

\begin{figure}[h]
  \centering
  \includegraphics[width=\linewidth]{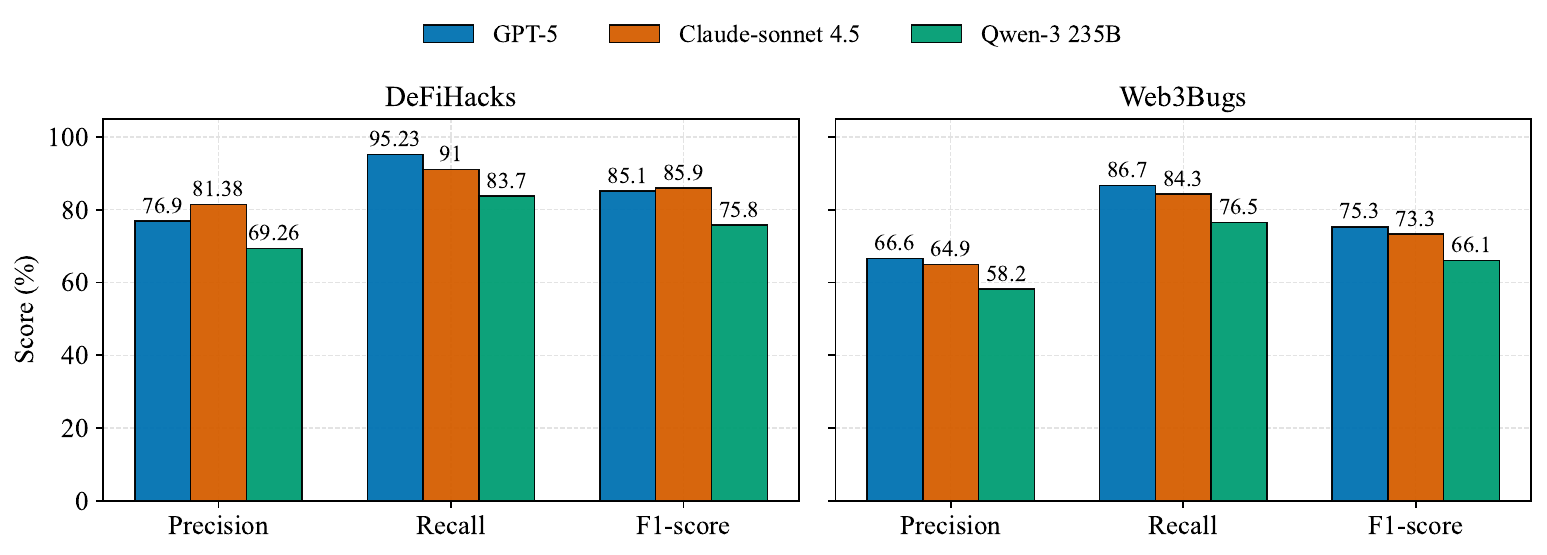}
  \caption{Comparison of the effectiveness of different LLMs on different datasets.}
  \label{fig:RQ2}
\end{figure}

\Cref{fig:RQ2} compares \tool's detection performance when powered by different LLM backends (GPT-5, Claude-sonnet 4.5, and Qwen-3 235B) on DeFiHacks and Web3Bugs, reporting precision, recall, and F1. 
Overall, \tool demonstrates strong model agnosticism: across different LLM backends, it maintains relatively stable detection performance. 
Nonetheless, we observe consistent differences among models. 
Compared with GPT-5, Claude-sonnet 4.5 tends to produce fewer false positives (i.e., higher precision) but slightly more misses (i.e., lower recall). The open-source Qwen-3 235B performs slightly worse than the two closed-source models overall, with lower precision and F1 scores.

\answer{2}{
Switching the backend LLM (GPT-5, Claude-sonnet 4.5, Qwen-3) yields consistently strong results. On DeFiHacks, precision is 69.26\%--81.38\% and recall is 83.7\%--95.23\%; on Web3Bugs, precision is 58.2\%--66.6\%. Claude trades fewer false positives for more misses, while Qwen-3 performs slightly worse. Overall, we believe that \tool is model-agnostic where its performance does not heavily rely on any specific LLMs.
}

\subsection{RQ3: Runtime performance and cost overhead}

In this section, we evaluate the running time and financial cost of \tool. To ensure a clear and accurate assessment, it is important to note that \tool consists of two distinct stages: Logic Miner and Logic Checker. We ignore the Miner stage in this evaluation because Miner is a one-time pre-processing task and is performed on a local infrastructure. For the Checker stage, we specifically measure the time and cost overhead associated with the three-phase procedure detailed in Section~\ref{sec:business_logic_vul_detect}: Invariant Induction, Consistency Verification, and Deviation Analysis. We measured the total usage of the three APIs on the DeFiHacks dataset through the corresponding model provider platform, OpenRouter.

\Cref{tab:model_comparison} presents the experimental results. In this table, Size (KLoc) represents the total lines of code in thousands, Total Time and Total Cost denote the cumulative execution time in seconds and monetary expenditure in USD, and Time / KLoc (s) and Cost / KLoc (\$) provide normalized metrics to assess efficiency and economic feasibility per thousand lines of code.
We counted the lines of code for 52 DeFi projects on the DeFiHacks platform, totaling 78,230 lines of code. The average total time spent was 10,222 seconds. The average time spent per project was 300 seconds. The average cost per thousand lines of code was \$0.08. Among them, the Claude Sonnet-4.5 offers lower latency, while the Qwen3 235B provides a more competitive price.

\begin{table}[htbp]
  \centering
  \caption{Efficiency and Economic Analysis of Contrastive Auditing Across LLM Backends.}
  \label{tab:model_comparison}
  \begin{tabular}{lccccc}
    \toprule
    \textbf{Model} & \textbf{Size (KLoc)} & \textbf{Time (s)} & \textbf{Cost (\$)} & \textbf{Time/KLoc(s)} & \textbf{Cost/KLoc(\$)} \\
    \midrule
    GPT-5 & 78.23 & 12392 & 7.71 & 158 & 0.0985 \\
    Claude Sonnet-4.5 & 78.23 & 9066 & 10.00 & 116 & 0.1280 \\
    Qwen3 235B & 78.23 & 9208 & 2.62 & 120 & 0.0300 \\
    \bottomrule
  \end{tabular}
\end{table}

\answer{3}{
\tool is fast and cost-effective, scanning each thousand lines of Solidity code in DeFiHacks datasets in just 131 seconds on average at a cost of \$0.08.
Qwen3 offers lower costs with latency comparable to Claude Sonnet-4.5, making it more suitable for large-scale scanning.}

\begin{figure}[h]
  \centering
  \includegraphics[width=0.8\linewidth]{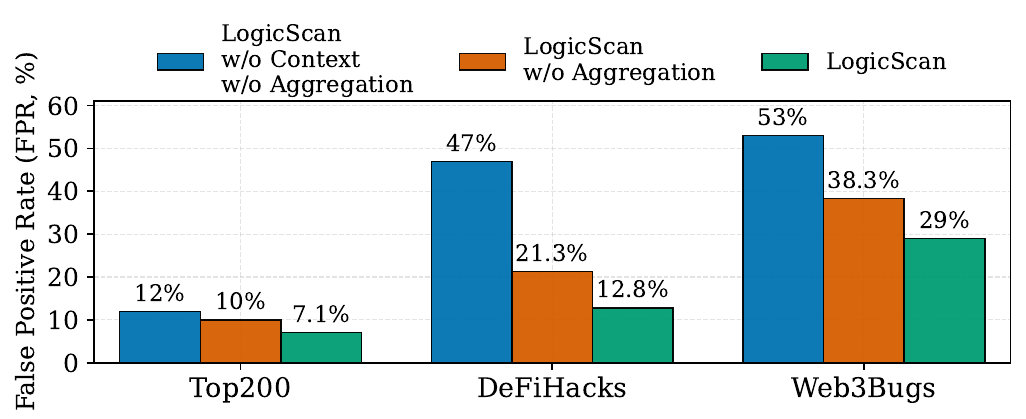}
  \caption{False positive rate (FPR) comparison across datasets.}
  \label{fig:RQ4}
\end{figure}

\subsection{RQ4: Ablation Study}

In RQ4, our goal is to evaluate whether introducing multi-dimensional code extraction and a noise-aware business-logic aggregation method proposed in Section~\ref{sec:llm_result_opt} can reduce the false positives produced by LLMs when analyzing smart contracts. To this end, we conduct an ablation study on \tool. We perform three groups of experiments on each of the three datasets mentioned above.

\textbf{\tool w/o Context w/o Aggregation.} We directly provide the retrieved template functions and the target function to the LLM in a single-round prompt, and ask the LLM to determine whether the target function is missing invariant-checking code that enforces the same logic as the corresponding business template function.

\textbf{\tool w/o Aggregation.} Building on \tool w/o Context \& Aggregation, we perform a second-stage extraction on the target function. Specifically, we extract the complete call-chain code of the target function, identify variables along the call chain, and further extract functions that perform write operations on those variables. We then feed the retrieved template functions, the target function, and the write-operation-related functions to the LLM via multi-round prompting.

\textbf{\tool.} Building on Context-Augmented, we aggregate the business logic of the retrieved template functions to obtain a common BSL. After the LLM read the business logic from the template code, we add an additional interaction round to supplement and reinforce its understanding of the generic business logic.

We compute the false positive rate (FPR) for each method on each dataset and report the results in Figure~\ref{fig:RQ4}. The false positive rate is defined as:
$\mathrm{FPR} = \frac{\mathrm{FP}}{\mathrm{FP} + \mathrm{TN}} $.
The FPR across all datasets exhibits a consistent downward trend. Specifically, the introduction of Multi-dimensional Code Extraction significantly reduces the FPR—dropping from 12\% to 10\% on the Top200 dataset, from 47\% to 21.3\% on DeFiHacks, and from 53\% to 38.3\% on Web3Bugs—demonstrating that providing the LLM with a complete call chain and write-operation context is particularly effective in mitigating false positives related to access control. Building on this, the full \tool tool, which incorporates business-logic aggregation, achieves the lowest FPR of 7.1\%, 12.8\%, and 29\% on the three datasets, respectively, indicating that noise-aware business-logic aggregation effectively filters out redundant and conflicting logic while reinforcing the model’s understanding of generic business invariants.

\answer{4}{
The ablation study results indicate that the integration of both multi-dimensional code extraction and noise-aware business-logic aggregation leads to a consistent and significant reduction in the false positive rate across all datasets. }

\subsection{Threats to validity}\label{sec:threats}

\textbf{Internal Threats.} 
\tool relies on large language models (LLMs) to generate functional descriptions, BSL, and invariant coverage analyses. 
Although we mitigate LLM instability through strict grammar constraints, syntactic and semantic validation, and bounded feedback loops, residual errors in semantic understanding may still occur, potentially leading to false positives or false negatives. 
In addition, our implementation depends on Slither for call-chain and variable-write extraction; any limitations of the underlying static analysis may affect context completeness and thus downstream reasoning. 
We reduce this risk via multi-dimensional code extraction, but cannot fully eliminate it.

\noindent\textbf{External Threats.} Our evaluation focuses on Ethereum-compatible DeFi contracts. While the datasets cover diverse and realistic scenarios, the results may not directly generalize to non-DeFi domains, non-EVM platforms, or ecosystems with substantially different execution or economic models. Applying \tool to such settings may require adapting the logic abstraction and template mining process. Additionally, although we evaluate multiple LLM backends, future models with different reasoning behaviors may affect performance.

\section{Related Work}\label{sec:related_work}

Early vulnerability detection mainly relied on static analysis. For example, slither \cite{feist2019slither}, Vandal \cite{brent2018vandal}, and smartcheck \cite{tikhomirov2018smartcheck}. They parse Solidity source code to construct intermediate representations (IR), combined with syntax parsers, to automatically identify potential security issues.
To handle cases without source code, tools such as Ethainter \cite{brent2020ethainter}, Zeus \cite{kalra2018zeus}, Securify \cite{tsankov2018securify}, and Mythril \cite{mythril2024} perform smart contract vulnerability detection based on bytecode. These tools decompile EVM bytecode and construct control flow or semantic models, using techniques such as symbolic execution or logical reasoning to identify security vulnerabilities in contracts.
In addition, there are detection methods designed for specific scenarios, such as reentrancy attacks\cite{zhang2022reentrancy,song2025silence}, arithmetic overflow issues \cite{murala2025enhancing}, third-party library dependency security\cite{liu2024using}, and access control issues\cite{ghaleb2023achecker,liu2022finding}.
Vulnerability detection methods based on formal analysis usually formalize contract behavior into logical models and use techniques such as model checking, symbolic execution, and abstract reasoning to automatically verify whether contracts meet specified safety and functionality specifications\cite{permenev2020verx,park2020end}.
However, static methods rely on detecting vulnerability patterns, making it difficult to represent business semantics. In complex scenarios, they often have a higher false positive rate.

Recently, with the emergence of generative artificial intelligence, LLMs can compensate for the above deficiencies through semantic understanding and collaborative reasoning. Multi-agent auditing systems enhance complex vulnerability recall and stability through role division and consensus\cite{wei2025advanced}.
For LLM-based detection, false positives and false negatives are common. Structured representations (such as logical semantic graphs) combined with CoT/ICL can strengthen vulnerability detection; at the same time, specialized models fine-tuned through pretraining/supervised learning can simultaneously improve consistency between “detection” and “explanation”\cite{yang2023automated,alam2024detection,mandana2025evullm,yu2025smart,sikder2025efficient,li2025scalm,wang2024smartinv}.
RAG can generate properties and link with verifiers to form a generation–verification closed loop\cite{liu2024propertygpt}.
Methods combining LLM-enhanced program analysis and symbolic execution (such as GPTScan\cite{sun2024gptscan}, NumScout\cite{chen2025numscout}, SymGPT\cite{xia2025symgpt}) can suppress LLM hallucination and improve the accuracy of vulnerability detection.

In the field of smart contract invariant extraction and generation, research has evolved across dynamic mining, static verification, and learning-based approaches. InvCon\cite{DBLP:conf/kbse/LiuL22} utilizes execution traces and the Daikon engine to mine likely invariants, while its successor, InvCon+\cite{DBLP:journals/tdsc/LiuZL25}, integrates Houdini-style static verification to ensure precise alignment with contract logic, primarily targeting standard protocols like ERC-20. Similarly, Trace2Inv\cite{DBLP:journals/pacmse/Chen0B0L24} deduces dynamic invariants for defense by mapping historical transaction trajectories to predefined security templates through dynamic taint analysis. Advancing into generative models, FLAMES\cite{eshghie2025flamesfinetuningllmssynthesize} leverages fine-tuning on extensive contract corpora and "Fill-in-the-Middle" tasks to directly synthesize compilable Solidity require statements, bypassing the limitations of template-based constraints. Unlike traditional tools that focus on successful transactions or internal code generation, RAVEN\cite{eshghie2025ravenminingdefensivepatterns} uniquely employs semantic clustering on reverted transactions on Ethereum to distill defensive invariant patterns that are already deployed and active in production environments.

Distinct from these works, \tool shifts the focus from intrinsic contract reasoning to external consensus-driven auditing. While current methods—static, dynamic, or LLM-based—rely on a target contract’s own code, they often miss fundamental logic gaps where critical constraints are entirely absent. By inducing invariants from a vast corpus of battle-tested protocols, \tool establishes a cross-protocol logic baseline. This enables the detection of deep-seated logic omissions and semantic inconsistencies that remain invisible when analyzing a contract in isolation.

\section{Conclusion}\label{sec:conclusion}
In this paper, we present \tool, an LLM-driven framework for detecting business logic vulnerabilities of smart contracts with business logic invariants mined from on-chain smart contracts.
A noise-aware contrastive auditing mechanism is designed to identify missing or weakly enforced invariants while mitigating LLMs' hallucination and the false positives caused.
\tool is evaluated on multiple real-world datasets, including DeFiHacks, Web3Bugs, and the Top-200 contracts.
The evaluation results indicate that \tool significantly outperforms the state-of-the-art approaches, achieving an F1 score of 85.2\% while maintaining a relatively low false positive rate.
In the future, we plan to combine \tool with dynamic analysis or formal verification techniques to further improve the detection accuracy of business logic vulnerabilities. 

\section*{Data Availability}
The artifact of this paper is available on Anonymous GitHub: \url{https://anonymous.4open.science/r/LogicScan-0660}.

\bibliographystyle{ACM-Reference-Format}
\bibliography{sample-base.bib}

\end{document}